\documentclass[conference]{IEEEtran}
\IEEEoverridecommandlockouts
\usepackage{cite}
\usepackage{amsmath,amssymb,amsfonts}
\usepackage{algorithm}
\usepackage{algorithmic}
\usepackage{graphicx}
\usepackage{textcomp}
\usepackage{xcolor}
\usepackage{booktabs}
\def\BibTeX{{\rm B\kern-.05em{\sc i\kern-.025em b}\kern-.08em
    T\kern-.1667em\lower.7ex\hbox{E}\kern-.125emX}}
\begin{document}

\title{FuncFooler: A Practical Black-box Attack Against Learning-based Binary Code Similarity Detection Methods\\

}

\author{\IEEEauthorblockN{1\textsuperscript{st} Lichen Jia}
\IEEEauthorblockA{\textit{SKLP} \\
\textit{Institute of Computing Technology, CAS}\\
Beijing, China \\
jialichen@ict.ac.cn}
\and
\IEEEauthorblockN{2\textsuperscript{nd} Bowen Tang \thanks{Lichen Jia and Bowen Tang contribute equally to the work.}}
\IEEEauthorblockA{\textit{SKLP} \\
\textit{Institute of Computing Technology, CAS}\\
Beijing, China \\
tangbowen@ict.ac.cn}
\and
\IEEEauthorblockN{3\textsuperscript{rd} Chenggang Wu}
\IEEEauthorblockA{\textit{SKLP} \\
\textit{Institute of Computing Technology, CAS}\\
Beijing, China \\
wucg@ict.ac.cn}
\and
\IEEEauthorblockN{4\textsuperscript{th} Zhe Wang \thanks{Zhe Wang is corresponding author.}}
\IEEEauthorblockA{\textit{SKLP} \\
\textit{Institute of Computing Technology, CAS}\\
Beijing, China \\
wangzhe12@ict.ac.cn}
\and
\IEEEauthorblockN{5\textsuperscript{th} Zihan Jiang}
\IEEEauthorblockA{\textit{SKLP} \\
\textit{Institute of Computing Technology, CAS}\\
Beijing, China \\
jiangzihan@ict.ac.cn}
\and
\IEEEauthorblockN{6\textsuperscript{th} Yuanming Lai}
\IEEEauthorblockA{\textit{SKLP} \\
\textit{Institute of Computing Technology, CAS}\\
Beijing, China \\
laiyuanming@ict.ac.cn}
\and
\IEEEauthorblockN{7\textsuperscript{th} Yan Kang}
\IEEEauthorblockA{\textit{SKLP} \\
\textit{Institute of Computing Technology, CAS}\\
Beijing, China \\
kangyan@ict.ac.cn}
\and
\IEEEauthorblockN{8\textsuperscript{th} Ning Liu}
\IEEEauthorblockA{\textit{School of Software} \\
\textit{Shandong University}\\
Jinan, China \\
liun21cs@sdu.edu.cn}
\and
\IEEEauthorblockN{9\textsuperscript{th} Jingfeng Zhang}
\IEEEauthorblockA{\textit{AIP} \\
\textit{RIKEN}\\
Tokyo, Japan \\
jingfeng.zhang@riken.jp}
}

\maketitle

\begin{abstract}
The binary code similarity detection (BCSD) method measures the similarity of two binary executable codes. Recently, the learning-based BCSD methods have achieved great success, outperforming traditional BCSD in detection accuracy and efficiency. However, the existing studies are rather sparse on the adversarial vulnerability of the learning-based BCSD methods, which cause hazards in security-related applications. To evaluate the adversarial robustness, this paper designs an efficient and black-box adversarial code generation algorithm, namely, FuncFooler. FuncFooler constrains the adversarial codes 1) to keep unchanged the program's control flow graph (CFG), and 2) to preserve the same semantic meaning.
Specifically, FuncFooler consecutively 1) determines vulnerable candidates in the malicious code, 2) chooses and inserts the adversarial instructions from the benign code, and 3) corrects the semantic side effect of the adversarial code to meet the constraints. 
Empirically, our FuncFooler can successfully attack the three learning-based BCSD models, including SAFE, Asm2Vec, and jTrans, which calls into question whether the learning-based BCSD is desirable.

\end{abstract}

\begin{IEEEkeywords}
binary code similarity detection, adversarial example
\end{IEEEkeywords}

\section{Introduction}

Binary code similarity detection (BCSD) acts as an major task in security-related applications such as vulnerabilities discovery~\cite{Esh,Tracelet}, malware detection~\cite{cesare2013control,ming2015memoized}, software plagiarism detection~\cite{luo2017semantics} and patch analysis~\cite{hu2016cross}. BCSD methods first disassemble the binary executable program, obtain the disassembled code, and cut the disassembled code in accordance with the code fragment granularity (function) to obtain code snippets, based on which the similarity of two executable programs are calculated. 

\begin{figure}[!htb]
    \setlength{\belowcaptionskip}{-5mm}
    
    \centering
    \includegraphics[scale=0.36]{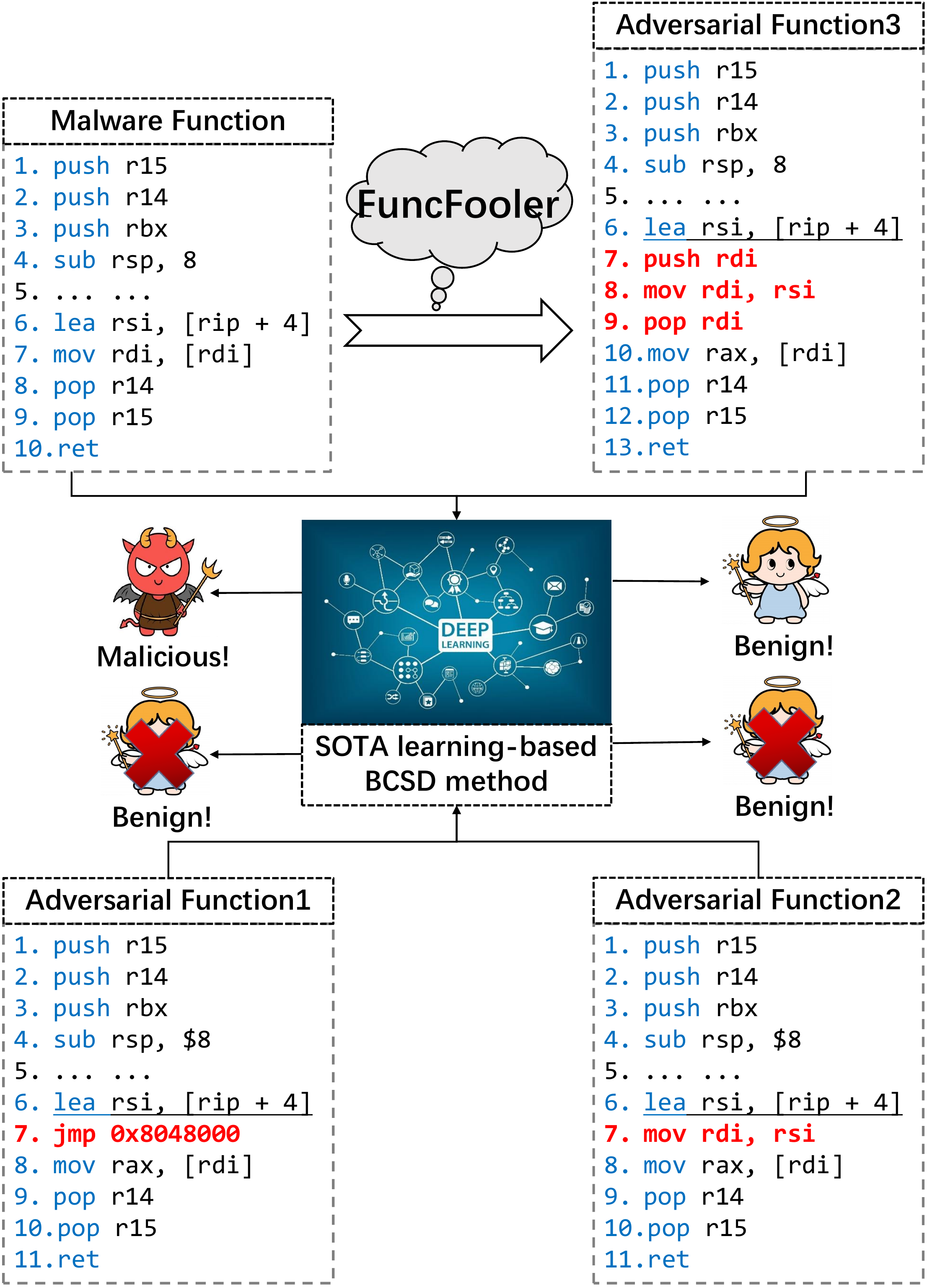}
    \caption{We pass the malware function and its three adversarial examples to the BCSD model, and the model outputs whether they are malicious or benign.}
    \label{fig:introduction}
\end{figure}

Traditional methods aim to analyze the feature of codes. For example, static methods~\cite{BinDiff,Binslayer} are based on graph isomorphism algorithms, which analyze the similarity of control flow graphs of two assembly code snippets; Dynamic methods~\cite{BinSim,blex} analyze the runtime behaviours of codes such as the patterns of system call or I/O pair to evaluate the similarity of code snippets. Yet, they are not satisfying for the BCSD problem due to the poor scalability and detection performance. Recently, with the vigorous development of learning-based methods in various domains, there has been an arising research trend to develop learning-based methods for BCSD tasks. These methods utilize deep learning~\cite{lecun2015deep} 
method to capture the semantic information of code snippets such as Asm2Vec~\cite{Asm2Vec}, SAFE~\cite{SAFE} and jTrans~\cite{jTrans}. Despite the success of learning-based methods on the BCSD tasks, previous works~\cite{chakraborty2021survey} have shown that deep learning methods are vulnerable to slight changes in input examples, especially the well-designed adversarial examples. Therefore, learning-based BCSD methods may suffer unaffordable costs due to the existence of adversarial codes. As  shown in Figure~\ref{fig:introduction}, malicious software with little changes can fool the learning-based BCSD system, violating the security protection systems. Therefore, it is of urgent need to revisit the current learning-based BCSD methods in terms of attacking. 

To the best of our knowledge, there are no recent works on the attack of learning-based BCSD methods. Although many kinds of adversarial attacks have appeared in the image\cite{image-survey2021} and text domains \cite{nlp-survey2020}, it is quite challenging to generate reasonable adversarial codes. Unlike images, the search space of adversarial codes can be discrete where the target assembled code must be chosen within a specific instruction set such as MIPS, Intel X86, making the search of adversarial codes difficult. Moreover, the designed adversarial examples should meet the semantic constraints, e.g, 1)  Adversarial examples in the BCSD  domain require their functionalities to be completely equivalent to the original examples. In addition, 2) the CFG of the adversarial example is the same as the CFG in the original example. The reason is many learning-based BCSD methods will sample instructions such as Asm2Vec according to the control flow graph. Changing the control flow graph of the program will affect the sampling results, thus affecting the prediction results of the model. Attacks on this sampling algorithm cannot account for the robustness of the model.

As shown in Figure~\ref{fig:introduction}, we get three adversarial examples by inserting instructions after the 6th instruction of the malware function. Although adversarial example1 can deceive the learning-based BCSD method, it cannot be regarded as an effective adversarial example due to the control flow graph being changed by the inserted jump instruction. adversarial example2 can deceive the learning-based BCSD method by inserting instruction \emph{mov rdi, rsi} after the instruction \emph{lea rsi, [rip+4]}, but it still cannot be regarded as an effective adversarial example because the functionality of adversarial example2 is different from malware function. The inserted mov instruction will cause the adversarial example2 to execute incorrectly. The adversarial example3 is generated by our work FuncFooler, which does not change the control flow graph and guarantees the functional equivalence with the malware function.


This paper presents FuncFooler, an adversarial code generation algorithm under the black box attack framework. It provides a general model-agnostic attack algorithm for current learning-based BCSD methods. To address the two challenges presented above, we designed three tasks in FuncFooler: 1) Determine vulnerable candidates. 2) Choose and insert adversarial instructions. 3) Correct the semantic side effect of the adversarial instructions. Task2 and task3 solve challenge1 and challenge2 respectively. Task1 can ensure that we minimize the number of adversarial instructions inserted when generating adversarial 
examples. We successfully applied this method to attack three state-of-the-art learning-based models in SPEC CPU 2006 and SPEC CPU 2017 benchmarks. On the adversarial examples, FuncFooler reduces the accuracy of all target models to below 5\% with only less than 13\% source instructions perturbed and lower than 1\% performance overhead. Overall, our study has the following contributions:

\begin{itemize}
\item We first introduce the challenges of adversarial attacks in binary code and redefine the concept of adversarial examples in binary code.

\item We propose a simple but practical attack method under the black-box setting, namely FuncFooler, which could efficiently generate reasonable adversarial examples under multiple constraints. 

\item We evaluate FuncFooler on three state-of-the-art deep learning models on SPEC CPU 2006\footnote{https://www.spec.org/cpu2006} and SPEC CPU 2017\footnote{https://www.spec.org/cpu2017}, and it can successfully attack the SOTA learning-based BCSD methods with a reasonable performance overhead.

\end{itemize}

\section{Problem Definition}
\subsection{Adversarial Attack Definition}
Before we define adversarial examples in the BCSD domain, we first introduce the problem definition of BCSD. The task of BCSD is to calculate the similarity of functions, One-to-many (OM) task calculates the similarity scores between the source function and all functions in the function pool, and sorts them according to the scores, it is the main application scenario as discussed in ~\cite{BCSD-survey,jTrans}. The tasks of the OM have the following definitions:



\subsubsection{Definition (Functions).} Function $f=\{I_{1}, I_{2}, I_{3}, ..., I_{n}\}$ is a set of ordered instructions, which is compiled from the source code function.  

\subsubsection{Definition (BCSD Tasks).} Given a source function $f_{src}$ and a pool of functions $FP = \{f_{1},f_{2},...,f_{n}\}$, the task of binary code similarity detection is to find the top-k functions with the highest similarity score from the function pool.







\subsubsection{Definition (adversarial example).} 
We redefine the adversarial example in the binary code domain as follows:

Given a function $f=\{I_{1}, I_{2}, I_{3}, ..., I_{n}\}$ with n instructions, an effective adversarial example $f_{adv}$ should meet the following requirements:

\begin{gather}
    sim(f, f_{adv}) \geq \epsilon \\
    functionality(f) == functionality(f_{adv}) \\ 
    f \notin rank(f_{adv}, FP, K) 
\end{gather}

The $sim(f, f_{adv})$ denotes the similarity between two functions, which is defined as $sim(f, f_{adv})= 1 - \frac{\|f_{adv}\| - \|f\|}{\|f\|}$. The adversarial example comes from the small disturbance to the source function, and their similarity should be greater than the threshold $\epsilon$. The $functionality(f)$ and $functionality(f_{adv})$ represents the functionality of $f$ and $f_{adv}$, and the two should be equivalent. The $rank(f_{adv}, FP, K)$ returns the top-k functions with the highest similarity score to the adversarial example $f_{adv}$ from FP. 

\section{Methodology}
\subsection{Overview}
\begin{figure}[h]
    \centering
    \includegraphics[scale=0.47]{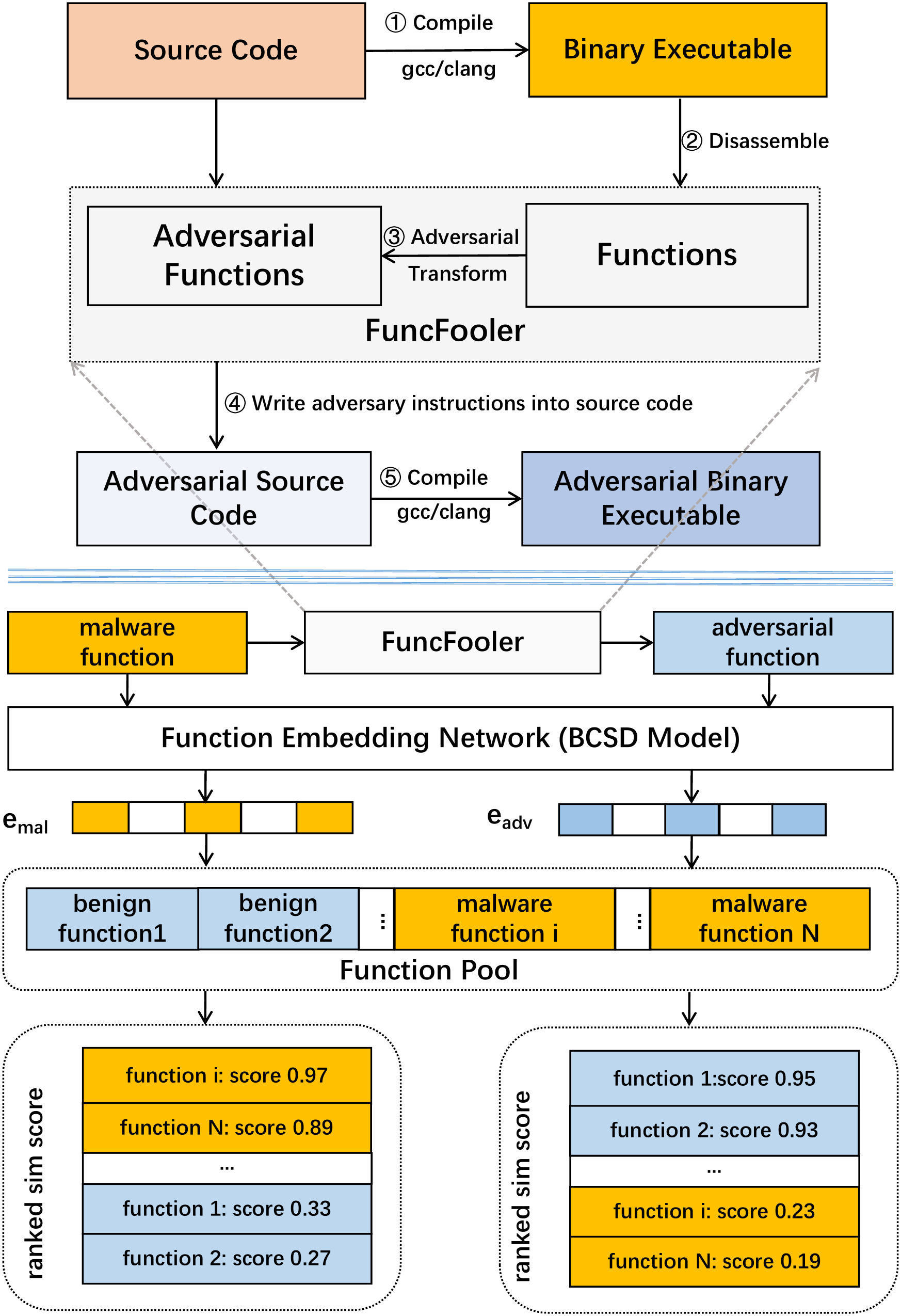}
    \caption{The workflow of how FuncFooler generates practical adversary binary executable in the real world. The embedding of malware function and adversarial example denote as $e_{mal}$ and $e_{adv}$,respectively}
    \label{fig:workflow}
\end{figure}

We will take the application of the learning-based BCSD method in the field of malware detection as an example, and introduce the role of FuncFooler in the field of learning-based BCSD malware detection. As shown in Figure~\ref{fig:workflow}, The input of FuncFooler is a malware function, and the output is an adversarial example. We pass the malware function and adversarial example to the learning-based BCSD model (Function Embedding NetWork) respectively, and the model will generate numerical vectors $e_{mal}$ and $e_{adv}$ respectively. These vectors are used by the model to calculate similarity scores with all functions in function pull, and are sorted in descending order of scores. From the similarity score, it can be seen that the malware function has a high similarity with other malicious functions (indicated in orange) with the same malicious behaviour in the function pool, and a low similarity with benign programs (indicated in blue). However, the adversarial example has high similarity with benign programs in the function pool, and low similarity with malicious functions.


Next, we will introduce how to use FuncFooler to get an adversarial binary executable. Suppose one person wants to hide features of all functions in a binary program executable from being detected by learning-based BCSD method, he will use FuncFooler as follows: 1) Use gcc/clang to compile the program source code into a binary executable program. 2) FuncFooler will disassemble the binary executable, and get the functions of the binary executable. 3) FuncFooler performs an adversarial transform on the functions by inserting instructions. We call these instructions adversarial instructions. 4) FuncFooler writes these adversarial instructions into the source code. 5) Use gcc/clang to compile the adversarial source code to get the adversarial binary executable.

\subsection{How to Generate adversarial example}
In this section, we introduce how FuncFooler generates an adversarial example. FuncFooler is a black-box attack and is not aware of the architecture, parameters, and training data of BCSD models. Algorithm~\ref{attack-algorithm} details the entire workflow of our attack including the adversarial generating algorithm. The adversarial example generation process mainly includes 3 tasks: 1) Determine vulnerable candidates[Algorithm~\ref{attack-algorithm}, line 1, 7]. 2) Choose and insert adversarial instructions [Algorithm~\ref{attack-algorithm}, line2-39]. 3) Correct the semantic side effect [Algorithm~\ref{attack-algorithm}, line33].

\begin{figure}

    \centering
    \includegraphics[scale=0.37]{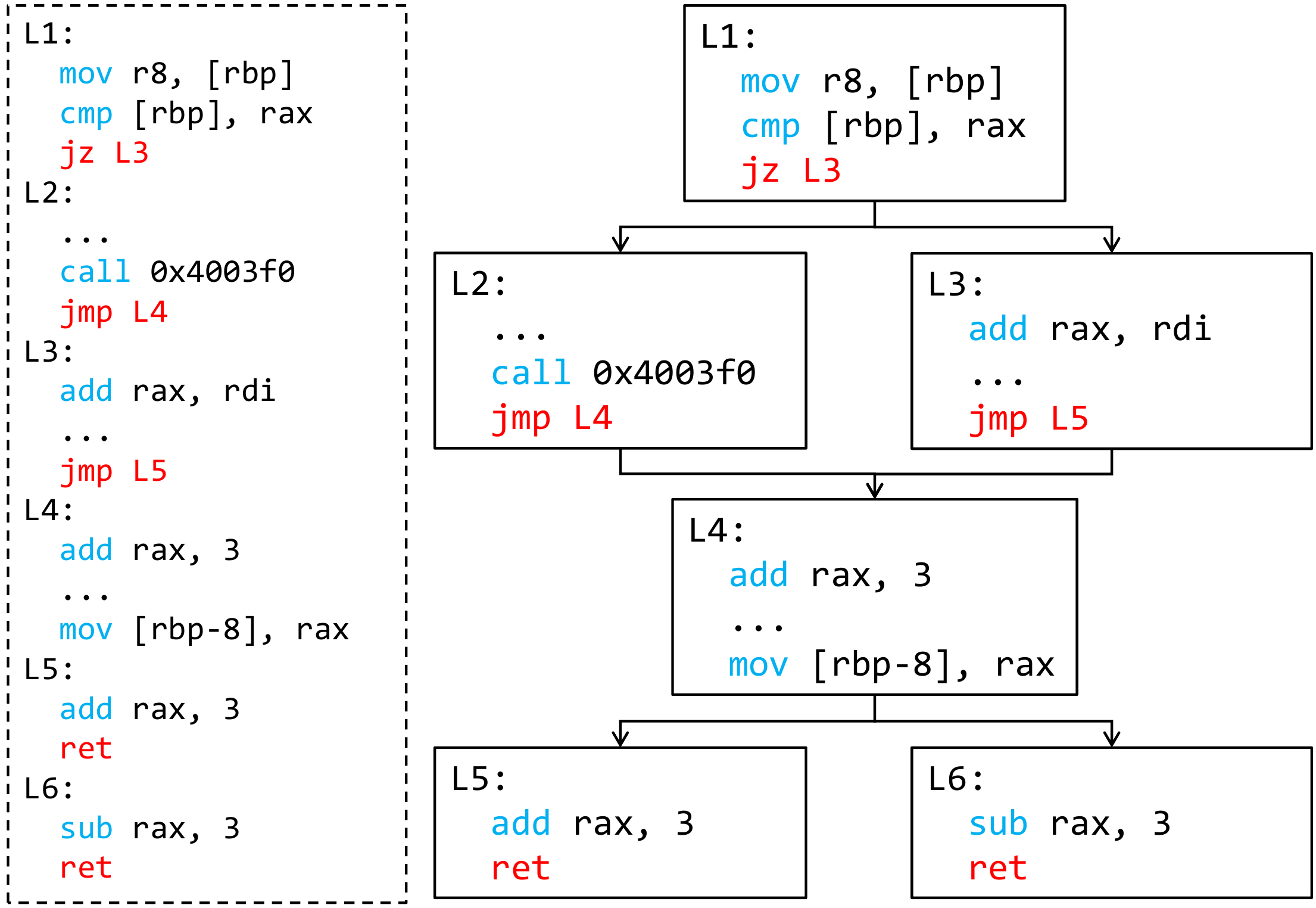}
    \caption{An example control flow graph of a binary function. The left part is linear layout assembly code with jump addresses, and the right part is the corresponding control-flow graph. L1 and L4 are Obligatory nodes.}
    \label{fig:Obligatory}
\end{figure}

\subsubsection{Task1: Determine Vulnerable Candidates.}
Learning-based BCSD usually adopts random sampling before training. The sampling process walks through the Control Flow Graph (CFG) randomly and samples the instruction sequences on each walking path. The \emph{Dominate Nodes} are the basic blocks located on all paths from the function entry to the exits, so the instructions of these nodes must be sampled regardless of choosing any walking path. Thus, attacking \emph{Dominate Nodes} will increase the success rate. Figure \ref{fig:Obligatory} depicts the CFG of an example function with 6 nodes. L1 and L4 are the \emph{Dominate Nodes} of the CFG. We will put all instructions in L1 and L4 to the Vulnerable Candidates Set. Here is the formal definition of our method: 

Given a function $f=\{I_{1}, I_{2}, ..., I_{n}\}$, we first construct its CFG with $Nodes=\{L_{1}, L_{2}, ..., L_{n}\}$, and apply the Tarjan Dominator Searching Algorithm~\cite{dominators} to collect the $Dominate\_Nodes=\{L^{dom}_{1}, L^{dom}_{2}, ..., L^{dom}_{n}\} \subseteq  Nodes$. Then, we pack all instructions in the $Dominate\_Nodes$ to the the candidate set $Intructions^{vul}=\{{I^{vul}_{1}, I^{vul}_{2}, ..., I^{vul}_{n}}\}$. 

\begin{algorithm}[!htb]

\caption{Adversarial Attack by FuncFooler}
\label{attack-algorithm}
\textbf{Input}: Function $f=\{I_{1},I_{2}, ..., I_{n}\}$, $rank(f, FP, k)$ rank function retrieve top-k similar most functions to $f$, $Rank^{gt}_{f}$ denotes their position of the ground truth in the list of retrieved functions. $sim(f, f_{adv})$ similarity calculation function, function similarity threshold $\epsilon$, function pool $FP = \{f_{1},f_{2},...,f_{n}\}$ \\
\textbf{Output}: Adversarial example $f_{adv}$ 
\begin{algorithmic}[1] 

\STATE $Intructions^{vul} \gets$ Find all vulnerable candidates in $f$

\STATE $I^{vul}_{k} \gets$ random sample one instruction from $intructions_{vul}$
\STATE $Inverses \gets rank(f, FP, -m)$ retrieve top m similar least functions to $f$  
\STATE Candidates $\gets \{\}$
\STATE
\FOR{$f' \in Inverses$}
    \STATE $Intructions^{adv} \gets$ Find all vulnerable candidates in $f'$
    
    \FOR{$I^{adv}_{j} \in Intructions^{adv}$}
        \STATE Initialization: $f^{adv} \gets$ $f$,  and copy $I^{adv}_{j}$ $N$ times ($N=Length(f)*0.2$) after $I^{vul}_{k}$
        \STATE Compute $Score_{I^{adv}_{j}}$ using equation~\ref{score}
    \ENDFOR

    
    \STATE Find All instructions $I^{adv}_{j} \in Intructions^{adv}$ with its $Score_{I^{adv}_{j}} \textgreater K$, add them to $Candidates$
    
    \STATE Filter out branch Instructions

    \IF{$len(Candidates) \geq num$} 
        \STATE Break
    \ENDIF
\ENDFOR
\STATE

\IF{Candidates is empty}
    \RETURN{None}
\ENDIF
\\

\STATE Initialization: $f^{adv} \gets f$
\WHILE{$f \in rank(f^{adv}, FP, k)$}
\STATE $I^{vul}_{k} \gets$ random sample from $Intructions^{vul}$
\STATE $I^{adv}_{j} \gets$ random sample from $(Candidates)$
\STATE Insert instruction $I^{adv}_{j}$ after $I^{vul}_{k}$
\STATE Correct the semantic side effect of $I^{adv}_{j}$
\ENDWHILE
\STATE
\IF{$sim(f, f^{adv}) \textgreater \epsilon$}
    \RETURN{$f^{adv}$}
\ENDIF

\RETURN{None}

\end{algorithmic}
\end{algorithm}

\subsubsection{Task2: Choose and Insert Adversarial Instructions.}
Adversarial instructions are those instructions that can make the model predict incorrectly, and an adversarial example can be obtained by inserting adversarial instructions in the vulnerable instructions candidate set $Intructions^{vul}$ obtained from task1. The functions with the least similarity in the pool are good sources, which are denoted as inverse functions $Inverses$. Therefore, we extract the \emph{Dominate Nodes} from the set of $f' \in Inverses$, and construct the adversarial instructions as $Intructions^{adv}$.


To determine the adversarial effect of an adversarial instruction $I^{adv}_{j}$, we randomly select a vulnerable instruction $I^{vul}_{k} \in Intructions_{vul}$, and copy $I^{adv}_{j}$ with $N$ times ($N=Length(f)*0.2$) after $I^{vul}_{k}$ to get the $f^{adv}=\{I_{1}, I_{2}, ..., I^{vul}_{k}, I^{adv}_{j},...,I^{adv}_{j}, I_{n}\}$. Then, we query the BCSD model and calculate the similarity score according to equation~\ref{score}. The $Rank^{gt}_{f^{adv}}$ and $Rank^{gt}_{f}$ represent the positions of their ground truth for source function $f$ and adversarial example $f^{adv}$ respectively.
\begin{equation}
  Score_{I^{adv}_{j}}  = Rank^{gt}_{f^{adv}}-Rank^{gt}_{f}
  \label{score}
\end{equation}
To get a reasonable attack example, we first evaluate the effect of all adversarial instructions $Intructions^{adv}$ in each inverse function $f' \in Inverses$, then find all instructions $I^{adv}_{j} \in Intructions^{adv}$ with its $Score_{I^{adv}_{j}} \textgreater K$, add them to $Candidates$ until the number of adversarial instructions in Candidates exceeds num (empirically, num=200).

Next, we randomly choose one adversarial instruction $I^{adv}_{j} \in Candidates$ and insert it after the instruction $I^{vul}_{k}$ we randomly in $Intructions^{vul}$. Correct the semantic side effects for adversarial instruction $I^{adv}_{j}$ to achieve the functional equivalence between $f$ and $f^{adv}$, where the correction method will be elaborated in the next section detailed. We repeat the insert and correct steps until we get a valid adversarial example. Finally, we check whether the adversarial examples satisfy the similarity constraints $sim(f, f^{adv}) \textgreater \epsilon$.

It should be noted that when selecting adversarial instructions, we have excluded control flow transfer instructions including \emph{call}, \emph{jmp}, \emph{ret} and all exception triggering instructions, including \emph{syscall} and \emph{INT}. The main reasons are as follows: 1) Inserting these instructions will greatly change the semantics, and the cost of correction is too huge. 2) Many learning-based BCSD methods have a preprocessing algorithm, which constructs a control flow graph for the function, and then samples the instruction sequence according to the control flow graph of the function. Inserting these instructions will affect the CFG, leading to wrong instruction sequences sampled by preprocessing algorithm. To rule out the reason for model inaccuracy caused by preprocessing algorithm, we need to keep the control flow graph of the program unchanged.

\subsubsection{Task3: Correct the Semantic Side Effect.}
Especially for programs, we must ensure the functional equivalence between the source function and the adversarial example. Inserting adversarial instructions in the source function can make the model predict incorrectly, but adversarial instructions will affect the state of registers and memory, resulting in program errors. We call this situation the side effect of adversarial instructions and divide them into the following three categories:

\begin{itemize}
    \item Modification to Common Registers, including general-purpose registers, floating-point registers, and SIMD registers;
    \item Modification to EFLAG Registers;
    \item Memory Corruption, including accessing illegally addresses or modifying the value of memory.
\end{itemize}

\begin{figure}[!h]
    \setlength{\belowcaptionskip}{-3mm} 
    \centering
    \includegraphics[scale=0.34]{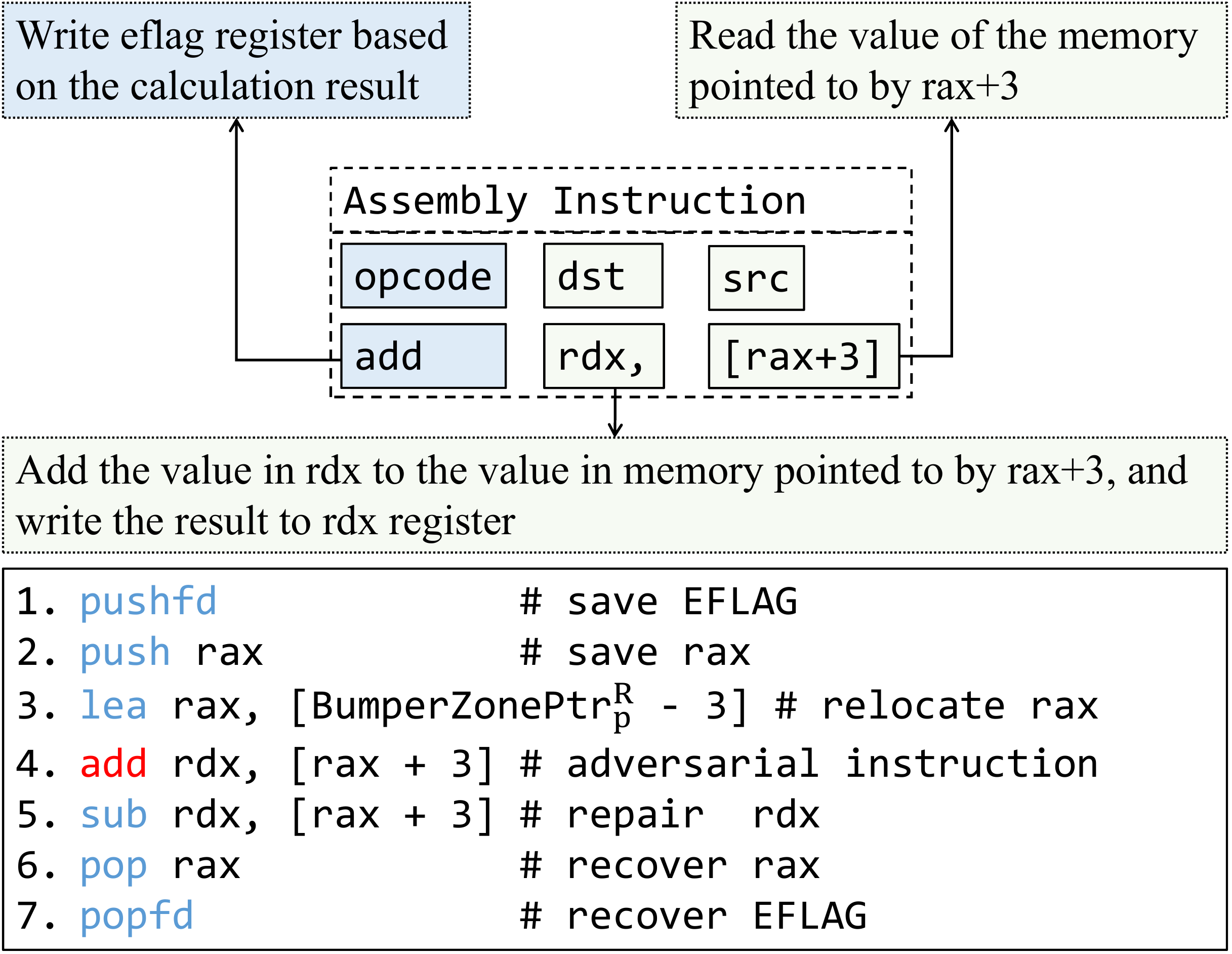}
    \caption{Side effects of adversarial instruction \emph{add rdx, [rax+3]} and the corresponding correction instruction. The instruction consists of one opcode add and two operands rdx, [rax+3]. Destination operand (dst) is rdx while source operand is [rax+3]. }
    \label{fig:assembly-code}
\end{figure}

As shown in Figure~\ref{fig:assembly-code}, assuming the adversarial instruction is an addition instruction as \emph{addl rdx, [rax+3]}, one of the source operands is \emph{rdx} register, the other is 1-byte value from memory with the address \emph{rax+3}, and the destination operand is \emph{rdx} register. The side effect of this instruction is a modification to the \emph{rdx} register, and the \emph{OF} and \emph{ZF} flags in the EFLAG register. In addition, if \emph{rax+3} points to an illegal memory area, the program will throw an exception. For these three cases, we have the following fixed solutions.

\subsubsection{Modification to Common Registers.} We have two fixed strategies: the first is that for some types of computation instructions, we can insert the inverse computation instructions after them. For example, in Figure~\ref{fig:assembly-code}, we can eliminate the side effect of 4th instruction on the \emph{rdx} register by adding a \emph{sub rdx, [rax+3]} instruction after it, but this method cannot handle load instructions, such as \emph{mov rax, [rax]}; the second is to save the destination register in memory in advance through the spill operation and restore it after the adversarial instruction. For example, for the instruction in Figure~\ref{fig:assembly-code}, the \emph{push rdx} instruction can be added before it, and the \emph{pop rdx} instruction can be added after it.

\subsubsection{Modification to EFLAG Registers.} Similar to the solution to the common register, we also have two fixed strategies for the EFLAG registers: the first is adding the inverse computation instruction after the adversarial instruction, this will repair the value of EFLAG automatically; the second is to protect EFLAG register via \emph{pushfd}/\emph{popfd} like spill/restore operation to the common register.

\subsubsection{Memory Corruption.} Our fixed strategy is to redirect the memory access address to a controllable memory area. To achieve the solution, we will insert multiple \emph{mmap} systems call into the initialization function \emph{\_libc\_start\_main()} in advance, so that the program will allocate multiple read-only memory areas $BumperZones^{R}$ and multiple readable and writable memory areas $BumperZone^{W}$ in the heap area within the process space during the initialization phase. The data in the $BumperZones^{R}$ is initialized to 0. We will store the address of each $BumperZone^R_i$ and $BumperZone^W_i$ in some global pointers $BumperZonePtr^R_i$ and $BumperZonePtr^W_i$ respectively. After that, we will insert the address redirection instruction before the adversarial instruction with  a memory corruption side effect. For example, for the instruction in Figure~\ref{fig:assembly-code}, we will insert the instruction \emph{mov rax, [$BumperZonePtr^R_p$ - 3]} in front of it, where $BumperZonePtr^R_p$ is a random address within a random p-th $BumperZonePtr^R$.

In Figure~\ref{fig:assembly-code}, to eliminate the memory corruption side effect of the 4th instruction, we insert 4 fix instructions 1, 3, 5, and 7. It should be noted that the fix instruction also has side effects and needs to be eliminated. The side effect of the 3rd instruction \emph{lea rax, [$BumperZonePtr^R_p$ - 3]} will modify the \emph{rax} register, thus we introduce another fix instructions 2 and 4 to eliminate it. And the 5th instruction \emph{sub rdx, [rax+3]} will also modify the EFLAG register, but no additional fix instruction needs to be added, because instructions 1 and 7 will eliminate it  together with the side effect of adversarial instruction.

In addition, for the benefit of patching on the assembly file generated by the compiler, except for a very small number of programs with embedding hand-written inline assembly code, all jump targets in the assembly file are represented in the form of literal labels, so any adversarial instruction inserted in any basic block will not affect the correctness of the control transfer.

\section{Experiment}

\subsection{Experimental Settings}
\begin{table*}[h]

  \centering
  \resizebox{\linewidth}{!}{
  
  \begin{tabular}{lccc|cccc|cccc|cccc}
    \toprule
    &              &      &              & \multicolumn{4}{c|}{SAFE} & \multicolumn{4}{c|}{Asm2Vec} & \multicolumn{4}{c}{jTrans} \\
                    & FN &  AFL & Overhead(\%) & OA & AA & \% IIR & CR & OA & AA & \% IIR & CR & OA & AA & \% IIR & CR \\
    \midrule
    SPEC CPU 2006  & 11120 & 79.15 & 0.61 & 100 & 4.48 & 10.95 & 8.72 & 100 & 8.34  & 12.11 & 7.57 & 100 & 6.24 & 5.99 & 15.65 \\  
    SPEC CPU 2017  & 16650 & 86.79 & 0.86 & 100 & 2.43 & 7.97  & 12.25 & 100 & 3.35 & 4.89 & 19.76 & 100 & 4.73 & 5.66 & 16.83 \\
    \bottomrule \\
  \end{tabular}
  }
  \caption{Attack effect of FuncFooler on three models with two function sets. FN refers to function numbers. AFL refers to average function length.}
  \label{table:adversary results}
  
\end{table*}

\subsubsection{Target Models.}
To verify the adversarial effect and performance overhead of FuncFooler, we select three SOTA learning-based BCSD works as our attack targets: a) SAFE~\cite{SAFE} employing a self-attentive RNN architecture to generate the embedding for a function; b) Asm2Vec~\cite{Asm2Vec} applying the PV-DM model to learn the abstract of the function; c) jTrans~\cite{jTrans} embedding the control flow information to the Transformer-based language models for function embeddings. SAFE\footnote{https://github.com/gadiluna/SAFE} and jTrans\footnote{https://github.com/vul337/jTrans/} offer the source code and out-of-box model trained by their custom training set. We re-implement asm2vec with default parameter settings since the closed source ecology.

\subsubsection{Binary Code Benchmark.}
Same as previous works~\cite{ren2021unleashing,zhang2022one,tsoupidi2020constraint}, we adopt SPEC CPU 2006 and SPEC CPU 2017 as the validation set. On the trained models, we construct the function pool through all the functions in the validation set. Then, we leverage FuncFooler to generate the adversarial example for each one of them and evaluate their perturbation magnitude and performance overhead. The original 2006 and 2017 SPEC CPU benchmarks contain 36 and 47 programs respectively. FuncFooler relies on WLLVM~\cite{wllvm} to convert the source code from C/C++/Fortran to assembly code. However, WLLVM does not support programs with embedding inline assembly codes and has some implementation bugs. We need to filter out the programs with compiling and linking errors. Finally, we get 19 programs with a total of 11120 functions on SPEC CPU 2006 and 28 programs with a total of 16650 functions on SPEC CPU 2017. Through these two function sets, we construct two function pools.


\subsubsection{Evaluation Metrics.} We perform a top-K (K=5) attack on SOTA learning-based BCSD methods. The retrieval performance can be evaluated using the following matrix, OA (Original Accuracy), AA (After-attack Accuracy), IIR (Inserted Instructions Ratio) and CR (Compressed Ratio):

\begin{footnotesize}
\begin{gather}
    OA = \frac{1}{|FP|}\sum_{f_{i} \in FP}I(Rank^{gt}_{f{i}} \leq K) \label{equation:OA} \\
    AA = \frac{1}{|AP|}\sum_{f^{adv}_{i} \in AP}I(Rank^{gt}_{f_{i}} \textgreater k) \label{equation:AA} \\
    IIR = \frac{\sum_{f_{i} \in FP}length(f^{adv}_{i}) - length(f_{i})}{\sum_{f_{i} \in FP}length(f_{i})} \label{equation:IIR} \\
    CR  = \frac{OA-AA}{IIR} \label{equation:CR} 
\end{gather}
\end{footnotesize}

The indicator function is defined as below:
\begin{footnotesize}
\begin{equation}
    I(x)= 
\left\{
    \begin{array}{lc}
        0, x = False \\
        1, x = True \\
    \end{array}
\right.
\end{equation}
\end{footnotesize}

We use OA to represent the original accuracy of the model. AA represents the recognition rate of the model against the adversarial samples. IIR stands for the percentage of instructions introduced into the original samples in order to generate adversarial examples. CR indicates how sensitive the models are in terms of adversarial instructions.

\subsubsection{Implementation Details.} Since inserting too many instructions not only affects the performance of the program but also increases the difficulty of correcting the semantic side effect, so we set $\epsilon$ to 0.8. For convenience, we encapsulate all target models into library modules and contract a common query interface to serve the subsequent adversarial generation algorithm. All experiments are performed on a server running the Ubuntu 16.04.7 operating system with Intel(R) Xeon(R) Gold 6132 CPU @ 2.60GHz, 256 GB RAM and 8 Nvidia Tesla V100 GPUs.

\subsection{Results}
We evaluate FuncFooler from three aspects: performance overhead, attack effect, and CR.

\subsubsection{Performance overhead.} As shown in Table~\ref{table:adversary results}, we evaluate the performance overhead of the adversarial examples generated by FuncFooler. FuncFooler has a greater advantage on the runtime overhead of the adversarial example, which matters to the practicability of the attack. In some scenarios, BCSD is exploited by reverse engineering, and the attack algorithm becomes a shield to protect the intellectual property of the code. As shown in Table~\ref{table:adversary results}, the average overhead of FuncFooler is less than 1\%.

\subsubsection{Attack effect.} Table~\ref{table:adversary results} summarizes the attacking result of FuncFooler to 3 target models on SPEC CPU 2006/2017 benchmarks. Overall, we conclude that it reduced the accuracy of all models from 100\% to below 9\%, at the cost of inserting less than 13\% instructions. However other non-intelligent attacking methods based on program transformation (obfuscation), such as the popular tool OLLVM~\cite{ollvm} and BinTuner~\cite{ren2021unleashing}, can only reduce the accuracy to about 20\%. Therefore, in terms of attacking success rate, FuncFooler is the performer. 

\subsubsection{CR analysis.} As we can see from Table~\ref{table:adversary results}, jTrans has a higher average CR than the other two models on SPEC CPU 2006 benchmarks. This is because jTrans utilizes a more complex transformer network Bert\cite{devlin2018bert} to add the precise position information of the instruction stream into the function embedding. More network parameters make it less robust.


\subsection{Transferability}
We examine the transferability of adversarial examples, i.e. whether adversarial examples based on one model can also fool another model. For this, we collect the adversary example from SPEC CPU 2006 and SPEC CPU 2017 that are wrongly predicted by one target model and then measured the prediction accuracy of them against another target model. 

\begin{table}[!htb]

\resizebox{\linewidth}{!}{
    \begin{tabular}{lllll}
        \toprule  
       & & Asm2Vec & SAFE & jTrans             \\
        \midrule
              & ASM2CEC & ---    & 47.28  & 46.55\\
 SPEC CPU 2006 &       SAFE    & 42.59   & ---   & 55.36\\
               & jTrans  & 58.61    & 57.13  & ---\\
        
        \midrule
                      & ASM2CEC & ---    & 46.27  & 46.55\\
 SPEC CPU 2017 &       SAFE    & 43.37   & ---   & 55.7\\
               & jTrans  & 59.7    & 56.2  & ---\\
        
        \bottomrule \\
    \end{tabular}
  }
  \caption{Transferability of adversarial examples on SAFE, Asm2Vec, and jTrans.}
  \label{transferability}
  \centering
\end{table}

As we can see from the results in Table~\ref{transferability}, adversarial examples we generate on one model still have attack effects on other models. Moreover, the adversarial examples generated based on the model jTrans show higher transferability.

\subsection{Adversarial Training}
\noindent Our work casts insights on how to better improve the learning-based BCSD models through these adversarial examples. We conduct a preliminary experiment on adversarial training, by feeding the models both the original data and the adversarial examples, to see whether the original models can gain more robustness. We collect some adversarial examples crafted from the SPEC CPU 2006/2017 that fooled Asm2Vec and add them to the original training set. Then, we apply Asm2Vec re-trained by the expanded training set on the remaining adversarial examples. 

\begin{table}[!htb]

\resizebox{\linewidth}{!}{
    \begin{tabular}{lcc|cc}
        \toprule
        & \multicolumn{2}{c|}{SPEC CPU 2006} & \multicolumn{2}{c}{SPEC CPU 2017} \\
        & AA & IIR & AA & IIR \\
        \midrule
        Original          & 8.34  & 12.11 & 3.35 &  4.89 \\
        $\textbf{+ Adv. Training}$ & \textbf{15.12} & \textbf{16.53} & \textbf{13.82} & \textbf{8.73}  \\
        \bottomrule \\
    \end{tabular}
   } 
    \caption{Comparison of the AA and IIR of original training (“Original”) and adversarial training (“+ Adv.Train”) of Asm2Vec Model.}
    \label{adv-traing}
    \centering
\end{table}

As Table~\ref{adv-traing} shows, both AA and IIR of Asm2Vec are improved by 50\% to 100\% on both benchmarks. This reveals that one of the potencies of our attack system can enhance the robustness of a model to future attacks by training it with the generated adversarial examples. However, after expanding the training set, it is not enough to defend against our attacks completely. We argue that new models and training methods need to be introduced to substantially improve the robustness of learning-based BSCD methods.

\subsection{Ablation Study}
In this subsection, we conduct a series of ablation experiments to study the effectiveness of task1 and task2. We do not perform an ablation study on task3 because task3 ensures that the adversarial example is semantic equivalent to the source function, and removing task3 causes the adversarial example to execute incorrectly.

\subsubsection{Determine Vulnerable Candidates.}
\noindent To verify the effectiveness of task1, We keep task2 and task3 the same as the original algorithm and maintain the $sim(f, f_{adv}) \textgreater \epsilon$. Then, we remove task1 and regard all instructions in the source function as vulnerable candidates, which is denoted as Random in Table~\ref{ablation-task1}. We choose Asm2Vec as our target model and measure AA on FuncFooler and Random. 

\begin{table}[h]

\resizebox{\linewidth}{!}{
    \begin{tabular}{lc|c}
        \toprule
        & SPEC CPU 2006 & SPEC CPU 2017   \\
        \midrule
        $sim$ & $\geq$ 80\% & $\geq$ 80\%     \\
        \textbf{FuncFooler}     & \textbf{8.34} & \textbf{3.35} \\
        Random  & 83.2 & 84.7 \\
        \bottomrule \\
    \end{tabular}
    }
    \caption{Ablation study for task1. Comparison of the AA before and after removing the task1. }
    \label{ablation-task1}
    \centering
\end{table}

The results are shown in Table~\ref{ablation-task1}. After removing task1, it can be seen that the AA of the target model has increased to 83.2\% and 84.7\% respectively. This suggests that the attack is ineffective without task1 because dominant nodes are critical to the learning-based BCSD, which determine the core of adversarial positions that can be selected. This strategy also reduces the number of adversarial instructions to reach the limitation of similarity and runtime overhead.

\subsubsection{Choose and Insert Adversarial Instructions.}
\noindent Same as the previous experiment, we delete task2 from FuncFooler and choose random instructions as adversarial instructions, which is denoted as Random in Table~\ref{ablation-task2}. In this study, we maintain the $sim(f, f_{adv}) \textgreater \epsilon$ and choose Asm2Vec as our target model and measure AA on FuncFooler and Random. 

\begin{table}[h]

\resizebox{\linewidth}{!}{
    \begin{tabular}{lc|c}
        \toprule
        & SPEC CPU 2006 & SPEC CPU 2017   \\
        \midrule
        $sim$ & $\geq$ 80\% & $\geq$ 80\%     \\    
        \textbf{FuncFooler}    & \textbf{8.34} & \textbf{3.35} \\
        Random  & 93.2 & 89.7 \\
        \bottomrule \\
    \end{tabular}
  }  
    \caption{Ablation study for task2. Comparison of the AA before and after removing the task2.}
    \label{ablation-task2}
    \centering
\end{table}

The results are shown in Table~\ref{ablation-task2}, we can see that the AA of the target model has increased to 93.2\% and 89.7\%, respectively. This suggests the attack is ineffective without task2. 
Task2 enables FuncFooler to find the instruction that is most likely to change the model prediction result, making the attack easier to succeed in the case of limited perturbations.

\section{Related Work}
\subsection{Traditional BCSD Methods}
\noindent Traditional binary code similarity detection techniques include dynamic and static methods. Dynamic methods~\cite{blex, ming2012ibinhunt} inspect the invariants of input-output or intermediate values of the program at runtime to check the equivalence of binary programs. These dynamic analysis-based methods suffer from limited code coverage and don’t scale well. Static methods \cite{BinDiff, BinHunt} rely on the syntactical and structural information of binaries, especially control-flow structures (i.e., organization of basic blocks within a function) to perform the matching. Compared with dynamic methods, static methods are more efficient but achieve lower accuracy, because it only captures syntactical and structural information of binaries but neglects the semantics and relationship between instructions.

\subsection{Learning-Based BCSD Methods}
Learning-based binary code similarity detection techniques map code snippets(function, basic block) to a numerical vector, which represents the semantics of the code snippets. SAFE \cite{SAFE}, Asm2Vec \cite{Asm2Vec} and jTrans \cite{jTrans} generate embeddings for binary function. jTrans is the-state-of-art work that embeds the individual binary function into embedding based on a jump-aware Transformer-based model. Deepbindiff \cite{Deepbindiff} generates embedding for binary basic block while Gemini \cite{Gemini} generate embedding for attributed control-flow graph(ACFG) using graph neural network. The learning-based BCSD methods outperform traditional BCSD methods, but the neural network has its vulnerability, which makes the learning-based BCSD methods also vulnerable. Our work FuncFooler sheds light on the phenomenon of vulnerability in the field of learning-based BCSD.

\subsection{Adversarial Attacks in Image and Text Domain}
Adversarial attacks reveal the weakness of DNN by cheating it with adversary samples, which differ from the original ones with only slight perturbations. The studies on adversarial attacks in the image domain~\cite{FGSM} make the DNN models predict wrong by adding human-imperceptible perturbations to specific pixels. The adversarial attack in the text domain~\cite{Hotflip} modifies the characters, words, and sentences in the text so that the DNN model predicts errors without changing the semantics of the original text as much as possible.

In the field of binary code similarity detection, code snippets can be treated as special texts, such as functions as sentences and instructions as words. However, as we mentioned in the introduction, code has its unique constraints, which makes adversarial attacks against images and text inapplicable to the BCSD domain. Therefore, we redefine the concept of adversarial examples in the field of BCSD and propose FuncFooler, which can generate adversarial examples that meet the requirements.

\section{Conclusion}
\noindent In this paper, we propose FuncFooler, the first practical adversary attack against the state-of-the-art learning-based BCSD method under black-setting. FuncFooler firstly determines vulnerable candidates in the source function, then obtain the adversarial example by choosing and inserting the adversarial into vulnerable candidates. Finally, FuncFooler corrects the semantic side effect of the adversarial instructions. Extensive results demonstrate the effectiveness of our proposed framework FuncFooler at generating adversary functions. Therefore, we hope that the learning-based BCSD method can draw inspiration from this work and design a more robust BCSD model.




\begin{thebibliography}{00}
\bibitem{BinDiff} Dullien, Thomas, and Rolf Rolles. "Graph-based comparison of executable objects (english version)." Sstic 5.1 (2005): 3.

\bibitem{BinHunt} Gao, Debin, Michael K. Reiter, and Dawn Song. "Binhunt: Automatically finding semantic differences in binary programs." International Conference on Information and Communications Security. Springer, Berlin, Heidelberg, 2008.

\bibitem{ming2012ibinhunt} Ming, Jiang, Meng Pan, and Debin Gao. "iBinHunt: Binary hunting with inter-procedural control flow." International Conference on Information Security and Cryptology. Springer, Berlin, Heidelberg, 2012.

\bibitem{lecun2015deep} LeCun, Yann, Yoshua Bengio, and Geoffrey Hinton. "Deep learning." nature 521.7553 (2015): 436-444.

\bibitem{ginius} Feng, Qian, et al. "Scalable graph-based bug search for firmware images." Proceedings of the 2016 ACM SIGSAC Conference on Computer and Communications Security. 2016.

\bibitem{Esh} David, Yaniv, Nimrod Partush, and Eran Yahav. "Statistical similarity of binaries." Acm Sigplan Notices 51.6 (2016): 266-280.

\bibitem{blex} Egele, Manuel, et al. "Blanket execution: Dynamic similarity testing for program binaries and components." 23rd USENIX Security Symposium (USENIX Security 14). 2014.

\bibitem{Multi-MH} Pewny, Jannik, et al. "Cross-architecture bug search in binary executables." 2015 IEEE Symposium on Security and Privacy. IEEE, 2015.

\bibitem{Binclone} Farhadi, Mohammad Reza, et al. "Binclone: Detecting code clones in malware." 2014 Eighth International Conference on Software Security and Reliability (SERE). IEEE, 2014.

\bibitem{MutantX} Hu, Xin, et al. "{MutantX-S}: Scalable Malware Clustering Based on Static Features." 2013 USENIX Annual Technical Conference (USENIX ATC 13). 2013.

\bibitem{Binsign} Nouh, Lina, et al. "Binsign: fingerprinting binary functions to support automated analysis of code executables." IFIP International Conference on ICT Systems Security and Privacy Protection. Springer, Cham, 2017.

\bibitem{Kam1n0} Ding, Steven HH, Benjamin CM Fung, and Philippe Charland. "Kam1n0: Mapreduce-based assembly clone search for reverse engineering." Proceedings of the 22nd ACM SIGKDD international conference on knowledge discovery and data mining. 2016.

\bibitem{Tracelet} David, Yaniv, and Eran Yahav. "Tracelet-based code search in executables." Acm Sigplan Notices 49.6 (2014): 349-360.

\bibitem{Binsequence} Huang, He, Amr M. Youssef, and Mourad Debbabi. "Binsequence: Fast, accurate and scalable binary code reuse detection." Proceedings of the 2017 ACM on Asia Conference on Computer and Communications Security. 2017.

\bibitem{TEDEM} Pewny, Jannik, et al. "Leveraging semantic signatures for bug search in binary programs." Proceedings of the 30th Annual Computer Security Applications Conference. 2014.

\bibitem{XMATCH} Feng, Qian, et al. "Extracting conditional formulas for cross-platform bug search." Proceedings of the 2017 ACM on Asia Conference on Computer and Communications Security. 2017.

\bibitem{Binslayer} Bourquin, Martial, Andy King, and Edward Robbins. "Binslayer: accurate comparison of binary executables." Proceedings of the 2nd ACM SIGPLAN Program Protection and Reverse Engineering Workshop. 2013.

\bibitem{BinSim} Ming, Jiang, et al. "{BinSim}: Trace-based Semantic Binary Diffing via System Call Sliced Segment Equivalence Checking." 26th USENIX Security Symposium (USENIX Security 17). 2017.

\bibitem{discovRE} Eschweiler, Sebastian, Khaled Yakdan, and Elmar Gerhards-Padilla. "discovRE: Efficient Cross-Architecture Identification of Bugs in Binary Code." NDSS. Vol. 52. 2016.

\bibitem{Bingo} Chandramohan, Mahinthan, et al. "Bingo: Cross-architecture cross-os binary search." Proceedings of the 2016 24th ACM SIGSOFT International Symposium on Foundations of Software Engineering. 2016.

\bibitem{SAFE} Massarelli, Luca, et al. "Safe: Self-attentive function embeddings for binary similarity." International Conference on Detection of Intrusions and Malware, and Vulnerability Assessment. Springer, Cham, 2019.

\bibitem{Asm2Vec} Ding, Steven HH, Benjamin CM Fung, and Philippe Charland. "Asm2vec: Boosting static representation robustness for binary clone search against code obfuscation and compiler optimization." 2019 IEEE Symposium on Security and Privacy (SP). IEEE, 2019.

\bibitem{innereye} Zuo, Fei, et al. "Neural machine translation inspired binary code similarity comparison beyond function pairs." arXiv preprint arXiv:1808.04706 (2018).

\bibitem{jTrans} Wang, Hao, et al. "jTrans: Jump-Aware Transformer for Binary Code Similarity." arXiv preprint arXiv:2205.12713 (2022).

\bibitem{Gemini} Xu, Xiaojun, et al. "Neural network-based graph embedding for cross-platform binary code similarity detection." Proceedings of the 2017 ACM SIGSAC Conference on Computer and Communications Security. 2017.

\bibitem{gao2018vulseeker} Gao, Jian, et al. "Vulseeker: A semantic learning based vulnerability seeker for cross-platform binary." 2018 33rd IEEE/ACM International Conference on Automated Software Engineering (ASE). IEEE, 2018.

\bibitem{GraphEmb} Massarelli, Luca, et al. "Investigating graph embedding neural networks with unsupervised features extraction for binary analysis." Proceedings of the 2nd Workshop on Binary Analysis Research (BAR). 2019.

\bibitem{Ordermatters} Yu, Zeping, et al. "Order matters: semantic-aware neural networks for binary code similarity detection." Proceedings of the AAAI Conference on Artificial Intelligence. Vol. 34. No. 01. 2020.

\bibitem{XBA} Kim, Geunwoo, et al. "Improving cross-platform binary analysis using representation learning via graph alignment." Proceedings of the 31st ACM SIGSOFT International Symposium on Software Testing and Analysis. 2022.

\bibitem{Deepbindiff} Duan, Yue, et al. "Deepbindiff: Learning program-wide code representations for binary diffing." Network and Distributed System Security Symposium. 2020.

\bibitem{Codee} Yang, Jia, et al. "Codee: a tensor embedding scheme for binary code search." IEEE Transactions on Software Engineering (2021).

\bibitem{DeepFool} Moosavi-Dezfooli, Seyed-Mohsen, Alhussein Fawzi, and Pascal Frossard. "Deepfool: a simple and accurate method to fool deep neural networks." Proceedings of the IEEE conference on computer vision and pattern recognition. 2016.

\bibitem{FGSM} Goodfellow, Ian J., Jonathon Shlens, and Christian Szegedy. "Explaining and harnessing adversarial examples." arXiv preprint arXiv:1412.6572 (2014).

\bibitem{one-pixel} Su, Jiawei, Danilo Vasconcellos Vargas, and Kouichi Sakurai. "One pixel attack for fooling deep neural networks." IEEE Transactions on Evolutionary Computation 23.5 (2019): 828-841.

\bibitem{Hotflip} Ebrahimi, Javid, et al. "Hotflip: White-box adversarial examples for text classification." arXiv preprint arXiv:1712.06751 (2017).

\bibitem{SCPN} Iyyer, Mohit, et al. "Adversarial example generation with syntactically controlled paraphrase networks." arXiv preprint arXiv:1804.06059 (2018).

\bibitem{iAdvT-Text-and-iVAT-Text} Sato, Motoki, et al. "Interpretable adversarial perturbation in input embedding space for text." arXiv preprint arXiv:1805.02917 (2018).

\bibitem{luo2017semantics} Luo, Lannan, et al. "Semantics-based obfuscation-resilient binary code similarity comparison with applications to software and algorithm plagiarism detection." IEEE Transactions on Software Engineering 43.12 (2017): 1157-1177.

\bibitem{hu2016cross} Hu, Yikun, et al. "Cross-architecture binary semantics understanding via similar code comparison." 2016 IEEE 23rd International Conference on Software Analysis, Evolution, and Reengineering (SANER). Vol. 1. IEEE, 2016.

\bibitem{cesare2013control} Cesare, Silvio, Yang Xiang, and Wanlei Zhou. "Control flow-based malware variantdetection." IEEE Transactions on Dependable and Secure Computing 11.4 (2013): 307-317.

\bibitem{ming2015memoized} Ming, Jiang, Dongpeng Xu, and Dinghao Wu. "Memoized semantics-based binary diffing with application to malware lineage inference." IFIP International Information Security and Privacy Conference. Springer, Cham, 2015.

\bibitem{BCSD-survey} Haq, Irfan Ul, and Juan Caballero. "A survey of binary code similarity." ACM Computing Surveys (CSUR) 54.3 (2021): 1-38.

\bibitem{BCSD-survey2020} Silva, Samuel Henrique, and Peyman Najafirad. "Opportunities and challenges in deep learning adversarial robustness: A survey." arXiv preprint arXiv:2007.00753 (2020).

\bibitem{chakraborty2021survey} Chakraborty, Anirban, et al. "A survey on adversarial attacks and defences." CAAI Transactions on Intelligence Technology 6.1 (2021): 25-45.

\bibitem{nlp-survey2020} Zhang, Wei Emma, et al. "Adversarial attacks on deep-learning models in natural language processing: A survey." ACM Transactions on Intelligent Systems and Technology (TIST) 11.3 (2020): 1-41.

\bibitem{image-survey2021} Nowroozi, Ehsan, et al. "A survey of machine learning techniques in adversarial image forensics." Computers \& Security 100 (2021): 102092.

\bibitem{ren2021unleashing} Ren, Xiaolei, et al. "Unleashing the hidden power of compiler optimization on binary code difference: An empirical study." Proceedings of the 42nd ACM SIGPLAN International Conference on Programming Language Design and Implementation. 2021.

\bibitem{dominators} Christensen, Henrik Knakkegaard, and Gerth Støling Brodal. Algorithms for finding dominators in directed graphs. Diss. Aarhus Universitet, Datalogisk Institut, 2016.

\bibitem{wllvm} Tristan Ravitch. Whole Program LLVM, https://github.com/travitch/whole-program-llvm, 2013.

\bibitem{ollvm} Junod, Pascal, et al. "Obfuscator-LLVM--software protection for the masses." 2015 IEEE/ACM 1st International Workshop on Software Protection. IEEE, 2015.

\bibitem{binkit} Kim, Dongkwan, et al. "Revisiting binary code similarity analysis using interpretable feature engineering and lessons learned." IEEE Transactions on Software Engineering (2022).

\bibitem{zhang2022one} Zhang, Haotian, et al. "One size does not fit all: security hardening of MIPS embedded systems via static binary debloating for shared libraries." Proceedings of the 27th ACM International Conference on Architectural Support for Programming Languages and Operating Systems. 2022.

\bibitem{wang2017composite} Wang, Shuai, Pei Wang, and Dinghao Wu. "Composite software diversification." 2017 IEEE International Conference on Software Maintenance and Evolution (ICSME). IEEE, 2017.

\bibitem{tsoupidi2020constraint} Tsoupidi, Rodothea Myrsini, Roberto Castañeda Lozano, and Benoit Baudry. "Constraint-based software diversification for efficient mitigation of code-reuse attacks." International Conference on Principles and Practice of Constraint Programming. Springer, Cham, 2020.

\bibitem{devlin2018bert} Devlin, Jacob, et al. "Bert: Pre-training of deep bidirectional transformers for language understanding." arXiv preprint arXiv:1810.04805 (2018).

\bibitem{IMF-SIM} Wang, Shuai, and Dinghao Wu. "In-memory fuzzing for binary code similarity analysis." 2017 32nd IEEE/ACM International Conference on Automated Software Engineering (ASE). IEEE, 2017.

\end{thebibliography}
\end{document}